\documentclass{PoS}
\usepackage{bm}
\usepackage{amssymb,amsfonts,amsmath,amsbsy,mathtext,cite,enumerate,float, bm}
\title{From neutrino electromagnetic interactions to spin oscillations in transversal matter currents}

\ShortTitle{Neutrino spin oscillations in transversal matter currents}

\author{\speaker{Alexander Studenikin}\\
                Department of Theoretical Physics, Faculty of Physics, Lomonosov Moscow State University, Moscow 119991, Russia\\
        Joint Institute for Nuclear Research, Dubna 141980, Moscow Region, Russia\\
        E-mail: \email{studenik@srd.sinp.msu.ru}}

\abstract{Massive neutrinos have nonzero magnetic moments. It is known that the neutrino spin oscillations can be induced by the neutrino magnetic moment interaction with the transversal magnetic field. We show that the effect of neutrino spin oscillations can be engendered in the absence of the neutrino magnetic moment and without the  transversal magnetic field. From a regions derivation of the effective neutrino evolution Hamiltonian in moving matter it follows that spin oscillations can be engendered by the interaction with the transversal current of matter. The obtained general expression for the neutrino effective evolution Hamiltonian can be used for investigations of various types of neutrino spin oscillations in the transversal matter currents considered in the neutrino mass and flavour basis.}

\FullConference{Neutrino Oscillation Workshop\\
		4 - 11 September, 2016\\
		Otranto (Lecce, Italy)}

\begin{document}

\section{Introduction}

It is well known that massive neutrinos participate in electromagnetic interactions (see \cite{Giunti:2014ixa}
for a review). One of the most straightforward consequences of neutrino nonzero mass
is the prediction \cite{Fujikawa:1980yx} that neutrinos can have nonzero magnetic
moments. Studies of neutrino magnetic moments and the related phenomena attract a reasonable interest in literature. The values of neutrino magnetic moments are constrained in the terrestrial laboratory experiments and in the astrophysical considerations (see, for instance,\cite{Beda:2012zz} and \cite{Raffelt:1990pj}).

One of the most important phenomenon of nontrivial neutrino electromagnetic
interactions is the neutrino magnetic moment procession and spin oscillations in presence of external
electromagnetic fields. Within this scope the neutrino spin oscillations $\nu^{L}\Leftrightarrow \nu^{R}$ induced by the neutrino magnetic moment interaction with the transversal magnetic field ${\bf B}_{\perp}$ was first considered
in \cite{Cisneros:1970nq}. Then spin-flavor oscillations $\nu^{L}_{e}\Leftrightarrow \nu^{R}_{\mu}$ in ${\bf B}_{\perp}$ in vacuum were discussed
in \cite{Schechter:1981hw}, the importance of the matter effect was emphasized in \cite{Okun:1986hi}.
The effect of the resonant amplification of neutrino spin oscillations in ${\bf B}_{\perp}$ in the presence of matter was proposed in \cite{Akhmedov:1988uk,Lim:1987tk}, the impact of the longitudinal magnetic field ${\bf B}_{||}$ was discussed in \cite{Akhmedov:1988hd}. Recently, we have considered in details \cite{Fabbricatore:2016nec} mixing and oscillations of neutrino mass and flavour states in an arbitrary constant magnetic field that have  the transversal ${\bf B}_{\perp}$ and longitudinal ${\bf B}_{||}$ nonzero components.

In a series of our papers \cite{Egorov:1999ah,Lobanov:2001ar, Dvornikov:2002rs} we proposed to use the generalized Bargmann-Michel-Telegdi Lorentz invariant equation for description of the neutrino spin $S_{\mu}$
evolution in arbitrary constant electromagnetic fields and moving matter, also accounting for other general types of non-derivative neutrino interactions with external fields. In \cite{Studenikin:2004bu, Studenikin:2004tv}
we used the three-di\-men\-sio\-nal equation for description of the neutrino spin vector $\bf S $ procession when the particle is propagating through moving matter  in presence of an electromagnetic field $F_{\mu \nu}=({\bf E}, {\bf B})$,
\begin{equation}\label{S}
\frac{d{\bf S}}
{dt}=\frac{2\mu}{ \gamma} \Big[ {\bf S} \times ({\bf
B}_0+{\bf M}_0) \Big],
\end{equation}
where $\gamma = (1-\beta^2)^{-\frac{1}{2}}$, $\bm{\beta}$ is the neutrino velocity. For definiteness, consider an electron neutrino with a generic magnetic moment $\mu$ and mass $m$ and matter  composed of electrons. The magnetic field in the neutrino rest frame $\bf{B}_0$ is determined by the transversal
and longitudinal (with respect to the neutrino motion) magnetic and electric field components in the
laboratory frame,
\begin{equation}
{\bf  B}_0=\gamma\Big({\bf B}_{\perp} +{1 \over \gamma} {\bf
B}_{\parallel} + \sqrt{1-\gamma^{-2}} \Big[ {\bf E}_{\perp} \times
\frac{{\bm\beta}}{\beta} \Big]\Big).
\end{equation}
The matter term ${\bf M}_0$ in Eq. (\ref{S}) is also composed of the transversal ${\bf  M}{_{0_{\parallel}}}$
and longitudinal  ${\bf  M}_{0_{\perp}}$ parts ,
\begin{equation}\label{M_0}
{\bf M}_0=\bf {M}{_{0_{\parallel}}}+{\bf M}_{0_{\perp}},
\end{equation}
\begin{equation}\label{M_0_parallel_perp}
\begin{array}{c}
\displaystyle {\bf M}_{0_{\parallel}}=\gamma{\bm\beta}\rho^{(1)}_{e} \left(1-{{\bf v}_e
{\bm\beta} \over {1- {\gamma^{-2}}}} \right){n_{0} \over
\sqrt {1- v_{e}^{2}}}, \ \ \ \ {\bf M}_{0_{\perp}}=-\rho^{(1)}_{e}{\bf v}_{e_{\perp}}\frac{n_{0}}{\sqrt {1-
v_{e}^{2}}},
\end{array}
\end{equation}
where  $\rho^{(1)}_e={1 \over {2\sqrt{2}\mu }}{G}_F(1+4\sin^2 \theta _W)$, and $n_0=n_{e}\sqrt {1-v^{2}_{e}}$ is the invariant number density of matter given in the reference frame for which the total speed of
matter $v_e$ is zero.
In (\ref{M_0_parallel_perp}) we neglect possible effects of matter polarization.
For the evolution between two neutrino spin states
$\nu_{e}^{L}\Leftrightarrow\nu_{e}^{R}$ under the influence of
weak interations with the transversal matter current ${\bf M}_{0_{\perp}}$
we get \cite{Studenikin:2004bu, Studenikin:2004tv} (see also
\cite{Studenikin:2016iwq}) the following equation
\begin{equation}\label{2_evol_eq}
	i\frac{d}{dt} \begin{pmatrix}\nu_{e}^{L} \\ \nu_{e}^{R} \\  \end{pmatrix}=
\frac{\mu}{\gamma}
	\begin{pmatrix}
	M_{0\parallel} & M_{0\perp}   \\
	  M_{0\perp}  & -M_{0\parallel}  \\
		\end{pmatrix}
	\begin{pmatrix}\nu_{e}^{L} \\ \nu_{e}^{R} \\ \end{pmatrix}.
\end{equation}
For the neutrino spin oscillation probability we get
\begin{equation}\label{ver2}
P_{\nu_{e}^L \rightarrow \nu_{e}^R} (x)=\sin^{2} 2\theta_\textmd{eff}
\sin^{2}{\pi x \over L_\textmd{eff}},
\ \ \ \ sin^{2} 2\theta_\textmd{eff}={M_{0\perp}^2\over
{M_{0\parallel}^2 +M_{0\perp}^2}}, \ \ \
L_\textmd{eff}={2\pi \over {\mu M_0}}\gamma.
\end{equation}
From (\ref{ver2}) it follows that neutrino spin oscillations can be induced not only by the neutrino interaction with the magnetic field ${\bf B}_{\perp}$ but also by neutrino interactions with matter in the case when there is a transversal matter current (or matter polarization) \cite{Studenikin:2004bu, Studenikin:2004tv, Studenikin:2016iwq}. Note that the dependence on the neutrino magnetic moment $\mu$ cancels out in (\ref{2_evol_eq}), and in (\ref{ver2}) as well. The existence of the discussed effect of neutrino spin oscillations engendered by the transversal matter current and matter polarization and its importance for astrophysical applications have been confirmed in a series of recent papers \cite{Cirigliano:2014aoa, Volpe:2015rla, Kartavtsev:2015eva, Dobrynina:2016rwy}.

Here below we demonstrate a consistent derivation of the effect of the neutrino spin oscillations in the transversal matter currents based on the direct calculation of the spin evolution effective Hamiltonian.  We consider two flavour neutrinos with two possible helicities $\nu_{f}= (\nu_{e}^{+}, \nu_{e}^{-}, \nu_{\mu}^{+}, \nu_{\mu}^{-})^T$ in moving matter composed of neutrons. The neutrino interaction Lagrangian reads
 \begin{equation}\label{Lagr}
 \begin{array}{c}
 {\it L}_{int}= -f^{\mu}\bar{\nu} (x)\gamma_{\mu}\frac{1+\gamma _{5}}{2}\nu (x), \ \ \ f^{\mu}=-\frac{G_F}{\sqrt 2}n(1,\bf v),
 \end{array}
 \end{equation}
 where the matter potential $f^{\mu}$ depends on the velocity of matter
 ${\bf v}=(v_1 , v_2 , v_3)$ and on the neutron number density in the
 laboratory reference frame $n=\frac{n_0}{\sqrt{1-v^2}}$.
 Each of the flavour neutrinos is a superposition of the neutrino mass states,
\begin{equation}\label{transformations}
    \begin{array}{c}
  \nu_{e}^{\pm} =\nu_{1}^{\pm}\cos\theta+\nu_{2}^{\pm}\sin\theta,\ \ \ \ \
  \nu_{\mu}^{\pm}=-\nu_{1}^{\pm}\sin\theta+\nu_{2}^{\pm}\cos\theta.
\end{array}
\end{equation}
The neutrino evolution equation in the flavour basis is
\begin{equation}\label{schred_eq_fl}
  i\dfrac{d}{dt}\nu_{f}=\Big(H^{eff}_{0} + \Delta H^{eff}\Big) \nu_{f},
\end{equation}
where the first term $H^{eff}_{0}$ of effective Hamiltonian determines the neutrino evolution in nonmoving matter. The second term $\Delta H^{eff}$ accounts for the effect of matter motion and it can be expressed as
\begin{equation}\label{delta_H}
\Delta H^{eff}=
\begin{pmatrix}
\Delta^{++}_{ee} & \Delta^{+-}_{ee} &\Delta^{++}_{e\mu} & \Delta^{+-}_{e\mu}  \\
\Delta^{-+}_{ee} & \Delta^{--}_{ee}  & \Delta^{-+}_{e\mu}  & \Delta^{--}_{e\mu}  \\
\Delta^{++}_{\mu e} & \Delta^{+-}_{\mu e} & \Delta^{++}_{\mu \mu} & \Delta^{+-}_{\mu \mu} \\
\Delta^{-+}_{\mu e} & \Delta^{--}_{\mu e} & \Delta^{-+}_{\mu \mu} & \Delta^{--}_{\mu \mu}
\end{pmatrix},
\end{equation}
where
\begin{equation}\label{HB}
\Delta^{ss'}_{kl}=\langle	{\nu_{k}^{s}}|\Delta H^{SM}|{\nu_{l}^{s'}}\rangle, \ \ \ k,l = e,\mu, \ \ s,s'=\pm.
\end{equation}
From (\ref{Lagr}) it follows that
\begin{equation}\label{delta_H}
\begin{array}{c}
\Delta H^{SM}=\frac{G_F}{2\sqrt 2}n \big(1+\gamma _{5}\big) {\bf v} {\bm \gamma}, \ \ \
{\bf v} {\bm \gamma}= v_1 \gamma _1 + v_2 \gamma _2 + v_3 \gamma _3 .
\end{array}
\end{equation}
In evaluation of $\Delta^{ss'}_{kl}$ we have first introduced
the neutrino flavour states $\nu_{k}^{s}$ and $\nu_{l}^{s'}$ as superpositions of
the mass states $\nu_{1,2}^{\pm}$. Then, using the exact
free neutrino mass states spinors,
\begin{equation}	 \label{wave func}
\nu_{\alpha}^{s}=C_{\alpha}\sqrt{\frac{E_{\alpha}+
m_{\alpha}}{2E_{\alpha}}}\begin{pmatrix}u^{s}_{\alpha} \\ \frac{\bm{\sigma p_{\alpha}}}{E_{\alpha}+m_{\alpha}}u^{s}_{\alpha}
\end{pmatrix}e^{i\bm{p_{\alpha} x}}, \ \ \ \alpha=1,2,
\end{equation}
where the two-component spinors $u^s_{\alpha}$
\begin{equation}\label{u_s_plus}
	u^{s=1}_{\alpha}=\begin{pmatrix}1 \\ 0\end{pmatrix}, \ \ \ \
	u^{s=-1}_{\alpha}=\begin{pmatrix}0 \\ 1\end{pmatrix},
\end{equation}
define neutrino helicity states, we have performed calculations that are analogous to
those performed in \cite{Fabbricatore:2016nec}. The difference
in calculations is that here we consider not electromagnetic
neutrino interaction with a magnetic field but the neutrino weak
interaction with moving matter given by (\ref{delta_H}). For the
typical term $\Delta^{ss'}_{\alpha \alpha '}=
\langle	{\nu_{\alpha}^{s}}|\Delta H^{SM}|{\nu_{\alpha '}^{s'}}\rangle$, that
by fixing proper values of $\alpha, s, \alpha'$ and $s'$ can reproduces all
of the elements of the neutrino evolution Hamiltonian $\Delta H^{eff}$
that accounts for the effect of matter motion, we obtain,
\begin{equation}\label{Delta_ss_aa}
 \Delta^{ss'}_{\alpha \alpha '}=
\frac{G_F}{2\sqrt 2}\frac{n_0}{\sqrt {1-v^2}}\Big\{{u^{s}_{\alpha}}^{T}
\Big[{ \big(1-\sigma_3\big)v_{\parallel}+
\big({\gamma_{\alpha \alpha '}}^{-1}\sigma_1 +i
{\widetilde{\gamma}_{\alpha  \alpha '}}^{-1}\sigma_2
\big)v_{\perp}}\Big]
u^{s'}_{\alpha'}\Big\},
\end{equation}
where $v_{\parallel}$ and $v_{\perp}$ are the
longitudinal and transversal velocities of the matter current and
\begin{equation}
{\gamma_{\alpha \alpha '}}^{-1}=\frac{1}{2}\big(
\gamma_{\alpha}^{-1}+\gamma_{\alpha '}^{-1}\big), \ \ \
{\widetilde{\gamma}_{\alpha \alpha '}}^{-1}=\frac{1}{2}\big(
\gamma_{\alpha}^{-1}-\gamma_{\alpha '}^{-1}\big), \ \ \
\gamma_{\alpha}^{-1}=\frac{m_\alpha}{E_\alpha}.
\end{equation}
Recalling expressions for the Pauli matrixes,
\begin{equation}\label{Paul_matrix}
\sigma _3 =\begin{pmatrix}
	1 & 0  \\
	  0 & -1  \\
		\end{pmatrix}, \ \ \ \ \sigma _1 =\begin{pmatrix}
	0 & 1  \\
	  1 & 0  \\
		\end{pmatrix},\ \ \ \ \sigma _2 =i\begin{pmatrix}
	0 & -1  \\
	  1 & 0  \\
		\end{pmatrix},
\end{equation}
we get
\begin{equation}\label{Delta_ss_aa_final}
 \Delta^{ss'}_{\alpha \alpha '}=
\frac{G_F}{2\sqrt 2}\frac{n_0}{\sqrt {1-v^2}}\Bigg\{{u^{s}_{\alpha}}^{T}
\Bigg[{ \begin{pmatrix}
	0 & 0  \\
	  0  & 2  \\
		\end{pmatrix}v_{\parallel}
 + \begin{pmatrix}
	0 & \gamma_{\alpha}^{-1}  \\
	  \gamma_{\alpha '}^{-1}  & 0  \\
		\end{pmatrix}v_{\perp}}\Bigg]
u^{s'}_{\alpha'}\Bigg\},
\end{equation}
The obtained expression (\ref{Delta_ss_aa_final}) confirms our previous result 
\cite{Studenikin:2004bu, Studenikin:2004tv} (see also \cite{Studenikin:2016iwq})
 that there is the effect of the neutrino
spin conversion and corresponding spin oscillations engendered by the interaction with
the transversal current of matter.  The obtained general expression
(\ref{Delta_ss_aa_final}) can be used for investigations of various
types of neutrino spin oscillations in the transversal matter currents considered in the neutrino mass and flavour basis.
It is clear that the corresponding effect engendered
by the transversal polarization of matter can be treated in much the same way.

The author is thankful to Paolo Bernardini, Gianluigi Fogli and
Eligio Lisi for the invitation to attend the Neutrino Oscillation
Workshop and to all of the organizers for their kind hospitality in
Otranto. This work is supported by the Russian Basic Research Foundation
grants No. 16-02-01023 and 17-52-53133.

\end{document}